\documentclass[prl,twocolumn]{revtex4-1}
\usepackage[utf8]{inputenc}
\usepackage{amsmath}
\usepackage{amsfonts}
\usepackage{amssymb}
\usepackage{bm}
\usepackage{graphicx}
\usepackage{xcolor}
\usepackage{microtype}

\begin{document}
\title{Spin imbalance induced transverse magnetization in the Hofstadter-Hubbard model}
\author{Bernhard Irsigler, Jun-Hui Zheng, Mohsen Hafez-Torbati, and Walter Hofstetter}
\affiliation{Institut f\"ur Theoretische Physik, Goethe-Universit\"at, Frankfurt am Main, Germany}
\begin{abstract}
The fermionic, time-reversal invariant Hofstadter-Hubbard model with a population difference between the two spin states is investigated. In the strongly interacting regime, where the system can be described by an effective spin model, we find an exotic spin structure by means of classical Monte-Carlo calculations. Remarkably, this spin structure exhibits a transverse net magnetization perpendicular to the magnetization induced by the population imbalance. It is thus inherently different from canted antiferromagnetism. We further investigate effects of quantum fluctuations within the dynamical mean-field  approximation and obtain a rich phase diagram including ferromagnetic, anti-ferromagnetic, ferrimagnetic, and transverse magnetization phases.
\end{abstract}
\maketitle
Artificial gauge fields are at the heart of ongoing research in the field of cold atomic gases \cite{Gerbier2010,Dalibard2011,Goldman2014,Goldman2014b,Goldman2016b} as they act as base for intriguing quantum matter such as topological insulators and exotic quantum magnetism. The latter requires strong interactions between the particles which makes it challenging to investigate theoretically as well as experimentally \cite{Hofstetter2018,Rachel2018}. However, recent experiments with ultracold atomic gases  have shown magnetic correlations in the driven optical lattice \cite{Gorg2018}, spin frustration \cite{Struck2011}, and antiferromagnetism (AFM) below the superexchange temperature \cite{Mazurenko2017} and shed light on the capability of cold atoms to create exotic states of quantum magnetism. It is thus interesting to reach for new phases in this context.

Besides the Haldane model the Harper-Hofstadter model is one of the most commonly used theoretical systems to study artificial gauge fields. The Hofstadter model in its time-reversal invariant version \cite{Goldman2010} has been realized in experiments with ultracold bosons \cite{Aidelsburger2013}. Theoretically it has been intensively studied for the interacting case and shows various insulating phases \cite{Kumar2016} as well as exotic magnetism induced by artificial gauge fields for the spin-1/2 \cite{Cocks2012,Orth2013} and also the spin-1 \cite{Hafez-Torbati2018} case. Similar theoretical studies were performed for the Haldane-Hubbard model \cite{Vanhala2016}. Spin-imbalanced fermions, on the other hand, show versatile features ranging from canted AFM \cite{Brown2017} to phase separation in traps \cite{Snoek2011,Sotnikov2013a} and in the dimensional crossover \cite{Revelle2016a}.
\begin{figure}
\centering
\includegraphics[width=\columnwidth]{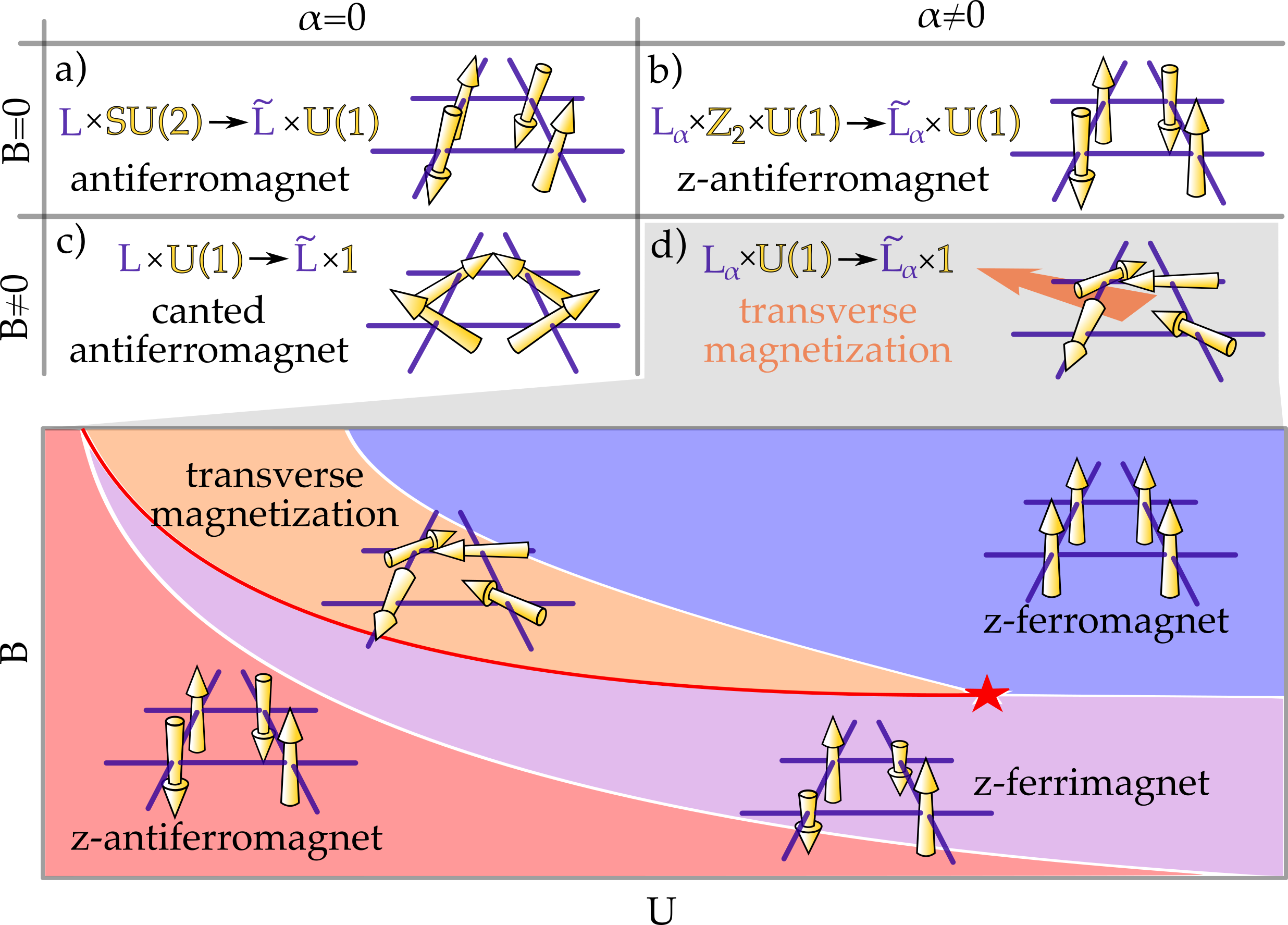}
\caption{Table of extensions to the Fermi-Hubbard model and its groundstates, considering artificial flux $\alpha$ and Zeeman field $B$. We represent lattice symmetries in purple and spin symmetries in yellow: a) the standard Hubbard model, b) the Hofstadter-Hubbard model, c) the spin-imbalanced Hubbard model, and d) the spin-imbalanced Hofstadter-Hubbard model including a schematic phase diagram.}
\label{table}
\end{figure}
In Fig.~\ref{table} we schematically depict the extensions of artificial gauge fields and spin-imbalance to the normal Hubbard model. Its groundstates differ strongly depending on whether an artificial flux $\alpha$ or a Zeeman field $B$ for spin-imbalance is applied. Fig.~\ref{table}a) shows the standard Hubbard model. Its Hamiltonian possesses the symmetry $L\times SU(2)$ where $L$ represents all the lattice symmetries of the system. The $SU(2)$ symmetry arises from rotational spin symmetry. After symmetry breaking, denoted by a black arrow, the AFM groundstate has the reduced lattice symmetry $\tilde{L}$ and a reduced spin symmetry $U(1)$. Long-range AFM correlations have been measured in a quantum gas experiment \cite{Mazurenko2017}. By applying a finite spin-dependent artificial flux one obtains the time-reversal invariant Hofstadter-Hubbard model shown in Fig.~\ref{table}b) which was subject of intensive theoretical studies \cite{Cocks2012,Orth2013,Kumar2016,Scheurer2015,Irsigler2019}. Here, the lattice symmetry $L_\alpha$ is reduced in comparison to $L$ since the size of the unit cell is now a mulitple of $1/\alpha$. The groundstate was found to be an AFM with the staggered magnetization pointing in the $z$-direction. Experiments come closer to measuring magnetic correlations in systems with artificial gauge fields as shown in Ref.~\cite{Gorg2018}. The spin-imbalanced Hubbard model shown in Fig.~\ref{table}c) yields canted AFM which has also been observed \cite{Brown2017}. We note that the systems in Fig.~\ref{table}b) and c) possess very different groundstates. In this work we investigate the groundstate of the spin-imbalanced Hofstadter-Hubbard model depicted in Fig.~\ref{table}d). It appears that the interplay of spin-imbalance and artificial gauge field induces a geometric frustration within the lattice similar to the Villain model \cite{Villain1977} which is also known as fully frustrated $XY$ model. This groundstate, in contrast to AFM or canted AFM then exhibits a finite transverse magnetization. The spin-imbalanced Hofstadter model has been studied recently in the non-interacting limit \cite{Yau2018} as well as for attractive interactions \cite{Iskin2015}. The Hamiltonian of the fermionic time-reversal invariant Hofstadter-Hubbard model reads

\begin{equation}
\begin{split}
\hat{H} =& -t\sum_{\bm{j}} \left[\hat{\bm{c}}_{\bm{j}+\hat{x}}^\dag \hat{\bm{c}}_{\bm{j}}
+ \hat{\bm{c}}_{\bm{j}+\hat{y}}^\dag e^{i\theta} \hat{\bm{c}}_{\bm{j}}
+ \text{h.c.}\right]\\
&+U\sum_{\bm{j}} \hat{n}_{\bm{j},\uparrow}\hat{n}_{\bm{j},\downarrow}.
\end{split}
\label{ham}
\end{equation}
Here, $\hat{\bm{c}}_{\bm{j}}^\dag=(\hat{c}_{\bm{j},\uparrow}^\dag,\hat{c}_{\bm{j},\downarrow}^\dag)$ is the fermionic creation operator, $\bm{j}=(x,y)$ is the lattice site vector, $\hat{x}=(1,0)$ and $\hat{y}=(1,0)$ are step vectors in the respective direction with the lattice constant set to unity, $t=1$ is the hopping energy, and $U$ is the interaction strength. When hopping in $y$-direction, the particle will pick up a phase $\theta=2\pi\alpha x \sigma^z$, where $\alpha$ is the plaquette flux, $\sigma^k$ is the $k$-th Pauli matrix with $k=x,y,z$, and $\hat{n}_{\bm{j},\sigma}=\hat{c}_{\bm{j},\sigma}^\dag\hat{c}_{\bm{j},\sigma}$  the density operator of the spin-$\sigma$ fermions.
For strong interactions the charge degrees of freedom freeze out and the Hamiltonian \eqref{ham} can be mapped onto an effective spin model \cite{Cocks2012,Orth2013}:
\begin{equation}
\label{spinHam}
\begin{split}
\hat{H}_\text{spin}&=
 J\sum_{\bm{j}} \left\{\hat{S}_{\bm{j}+\hat{x}}^x\hat{S}_{\bm{j}}^x
 +\hat{S}_{\bm{j}+\hat{x}}^y\hat{S}_{\bm{j}}^y
 +\hat{S}_{\bm{j}+\hat{x}}^z\hat{S}_{\bm{j}}^z\right\}\\
 &+J\sum_{\bm{j}}\left\{\cos(4\pi\alpha x)\left[\hat{S}_{\bm{j}+\hat{y}}^x\hat{S}_{\bm{j}}^x+\hat{S}_{\bm{j}+\hat{y}}^y\hat{S}_{\bm{j}}^y\right]\right.\\
 &\left.+\sin(4\pi\alpha x)\left[\hat{S}_{\bm{j}+\hat{y}}^y\hat{S}_{\bm{j}}^x-\hat{S}_{\bm{j}+\hat{y}}^x\hat{S}_{\bm{j}}^y\right]+\hat{S}_{\bm{j}+\hat{y}}^z\hat{S}_{\bm{j}}^z\right\}
 \end{split}
\end{equation}
where $J=t^2/U$ is the superexchange interaction energy and $\hat{S}^i_{\bm{j}}=\hat{\bm{c}}_{\bm{j}}^\dag \sigma^i \hat{\bm{c}}_{\bm{j}}$ is the spin operator. Note that for $\alpha=1/2$ this Hamiltonian simplifies to the two-dimensional Heisenberg Hamiltonian. In this work we will focus on the case $\alpha=1/4$. Other fluxes require larger unit cells and are not considered here. There are two ways to introduce a population spin-imbalance and thus break the time-reversal invariance: either introducing a Zeeman field term $-B\sum_{\bm{j}}\hat{S}^z_{\bm{j}}$ or allowing only states of the proper fixed magnetization $S^z=1/N_s\sum_{\bm{j}}\langle\hat{S}^z_{\bm{j}}\rangle$, where $N_s$ is the number of lattice sites. Assuming a product state in real space of the many-body system, the spin operators $\hat{S}^i_{\bm{j}}$ in Eq.~\eqref{spinHam} can be replaced by their respective expectation values $S^i_{\bm{j}}=\langle\hat{S}^i_{\bm{j}}\rangle$, such that $|\bm{S}_{\bm{j}}|=1$. This corresponds to a classical approximation of the quantum spin model Eq.~\eqref{spinHam}. We determine the groundstate of this approximated model using a classical annealing Monte-Carlo (CMC) algorithm. In order to fix the magnetization $S^z$ exactly we constrain the algorithm to states with the desired magnetization $S^z$. This is guaranteed by applying the following constraint during the CMC procedure similar to Ref.~\cite{Asselin2010}: starting with a random initial state with the correct $S^z$, a site $\bm{j}$ is randomly picked and its spin $S_{\bm{j}}$ is randomly flipped. A second site $\bm{j'}$ is randomly picked and the $z$-component $S^z_{\bm{j'}}$ is adjusted such that the correct magnetization $S^z$ is recovered if possible, otherwise it is a null-move. The remaining $S^x_{\bm{j'}}$- and $S^y_{\bm{j'}}$-components of the spin at site $\bm{j'}$ are chosen randomly obeying the normalization $|\bm{S}_{\bm{j'}}|=1$. Note, that this CMC procedure acts on three degrees of freedom instead of two in the normal, unconstrained CMC procedure.

\begin{figure}
\centering
\includegraphics[width=\columnwidth]{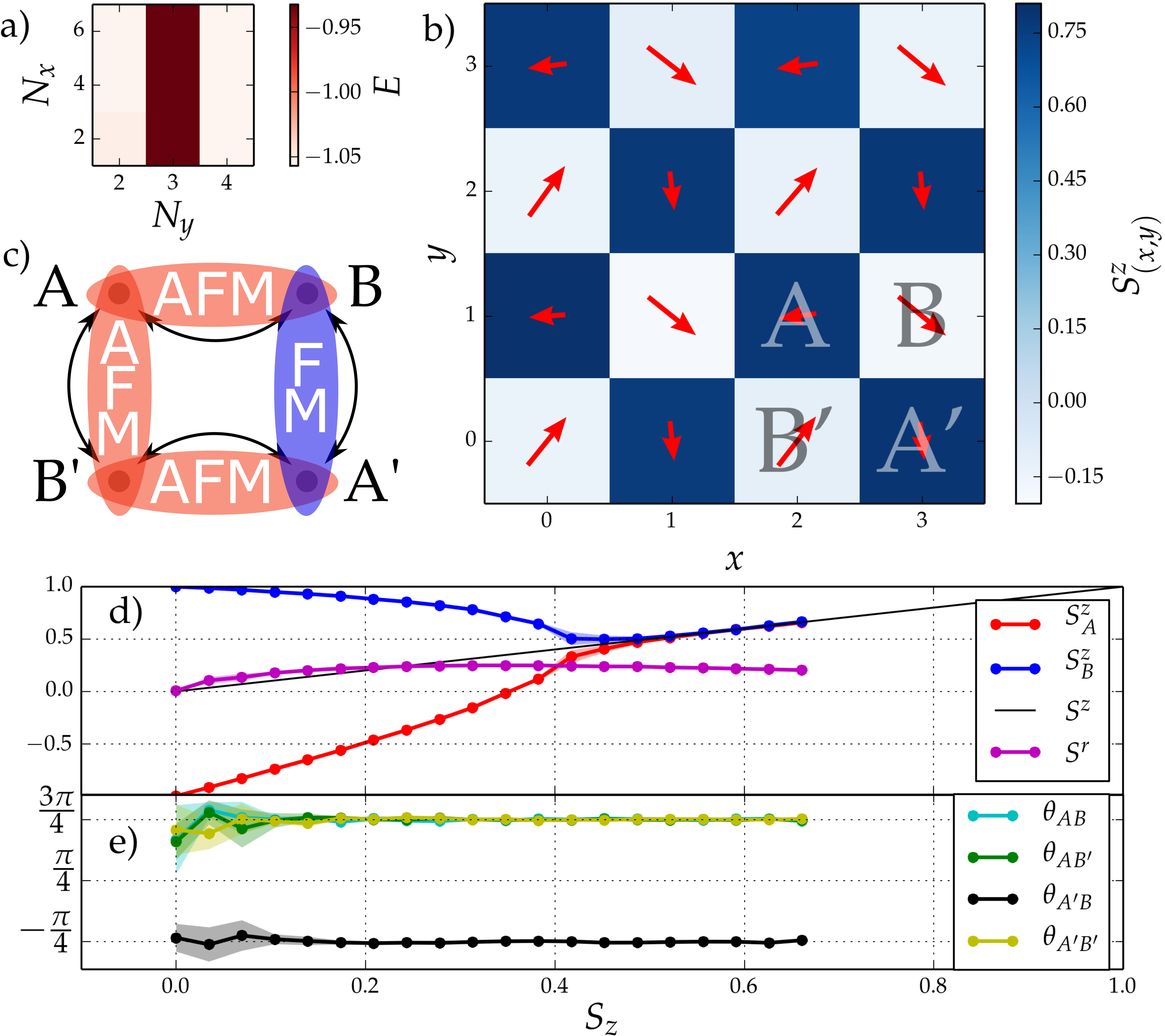}
\caption{Classical Monte-Carlo results: a) Groundstate energy $E$ of different unit cells of size $N_x\times N_y$ for $\alpha=1/4$, b) spin structure for $S^z=1/3$. Red arrows show spin orientations in the $S^x$-$S^y$-plane, the colormap represents the $S^z$-components. The labeling of the unit cell is shown in the right lower corner. c) Schematic of the preferred magnetic coupling between the spin projections onto the $S^x$-$S^y$-plane of the sites in the unit cell. d) $S^z$-components of the two sublattices $A$ and $B$ and transverse magnetization $S^r$. e) Angles between the spin projections of neighboring sites onto the $S^x$-$S^y$-plane.}
\label{result1_4}
\end{figure}

The Hamiltonian \eqref{spinHam} exhibits Heisenberg-type interactions in $x$-direction and $x$-dependent spin interactions in $y$-direction with a period of $n/2\alpha$ where $n$ is some integer. In the case $\alpha=1/4$ we find that the symmetry of the Hamiltonian requires the unit cell to have at least $N_x=2$ sites in $x$-direction. A priori we cannot make similar considerations for the number of sites $N_y$ in the $y$-direction since the symmetry will be spontaneously broken. In Fig.~\ref{result1_4}a) we show the groundstate energy per site for different sizes of the unit cell $(N_x,N_y)$ obtained from constrained CMC runs for $\alpha=1/4$ and $S^z=1/3$. We only show the lowest energy out of 100 of the respective CMC results. We observe that multiples of the 2$\times$2-unit cell yield the same energy which is lower compared to unit cells which are not multiples of 2$\times$2. We conclude that 2$\times$2 is the correct size of the unit cell.

In Fig.~\ref{result1_4}b) we show a spin configuration obtained from constrained CMC runs for $S^z=1/3$ and provide a labeling of the unit cell, where the letter (A,B) represents the magnitude of the $S^z$-component and the prime denotes different spin orientations in the $S^x$-$S^y$-plane. In Fig.~\ref{result1_4}d) we show the two $S^z$-components of the two sublattices $A$ and $B$, $S^z_A$ and $S^z_B$, respectively, as function of the total magnetization in $z$-direction $S^z$. They sum up to $S^z=S^z_A+S^z_B$ for every value of $S^z$. For $S_z=0$ we find $S^z_A=-1$ and $S^z_B=1$, respectively, corresponding to the AFM phase. For $S_z=1$ we find $S^z_A=S^z_B=1$, corresponding to the ferromagnetic (FM) phase. For finite magnetization $S^z\lesssim0.4$ their magnitude reduces forcing the spins into the $S^x$-$S^y$-plane due to $|S_{\bm{j}}|=1$. We observe that the spin-components in the $S^x$-$S^y$-plane do not cancel out as they do for instance in the canted anti-ferromagnetic phase \cite{Sotnikov2013a}. This leads to a finite total magnetization in the $S^x$-$S^y$-plane 
\begin{equation}
S^r=\frac{1}{N_xN_y}\sqrt{\left(\sum_{\bm{j}}S^x_{\bm{j}}\right)^2+\left(\sum_{\bm{j}}S^y_{\bm{j}}\right)^2},
\label{trans}
\end{equation}
shown as magenta line in Fig.~\ref{result1_4}d), which we call transverse magnetization phase (TM) as the systems responds to longitudinal magnetization $S^z$ with a transverse magnetization $S^r$ which can even exceed the magnitude of $S^z$ for $S^z\lesssim0.2$.
At $S^z\approx0.4$ we observe a phase transition at which both $S^z$-components become polarized $S^z_A=S^z_B$. Here, the transverse magnetization remains finite, but has to vanish at full polarization $S^z=1$. The regime of high magnetization $S^z>0.7$ is not well accessible with constrained CMC, since the number of null-moves increases drastically. The projections of the spin vectors of site $\bm{j}$ and site $\bm{j'}$ onto the $S^x$-$S^y$-plane form the angle $\theta_{\bm{jj'}}$. In Fig.~\ref{result1_4}e) we show these angles of all sites within the unit cell. At $S^z=0$ these angles are ill-defined due to AFM order in $S^z$-direction. For small values of $S^z\lesssim0.1$ the fluctuations of the angles is high, since the $S^x$-$S^y$-contribution to the energy is small, leading to fluctuations in the CMC procedure. For all remaining values of $S^z$ we find convergence for odd $x$ to an angle $\theta_{A',B}=-\pi/4$ along the $y$-direction and for even $x$ we find an angle $\theta_{A,B'}=3\pi/4$ along the $y$-direction. Along the $x$-direction we find solely the angles $\theta_{A,B}=\theta_{A',B'}=3\pi/4$. Note the sign convention of the angles, since all four angles have to add up to $2\pi n$. The angles can be explained from the Hamiltonian \eqref{spinHam} and are schematically shown in Fig.~\ref{result1_4}c). For $\alpha=1/4$ the sin-function vanishes and the coupling along the $y$-direction is FM for $x$ being odd and AFM for $x$ being even. This corresponds to an anti-ferromagnetic version of frustrated spins on a square lattice as first proposed by Ref.~\cite{Villain1977}. It is furthermore known as the fully frustrated $XY$ model and is of major interest, especially because of its two phase transitions being a chiral Ising-type transition and a Berizinskii-Kosterlitz-Thouless transition \cite{Teitel1982,Yosefin1985,Gabay1988,Ramirez1991,Olsson1995,Ozeki2003}.   It is important to recall that the present system, in contrast to the fully frustrated $XY$ model, consists of three-dimensional spins. The parallels can thus only be drawn for finite Zeeman field $B$ such that the spins acquire in-plane components. We think that the more general case of the uniformly frustrated $XY$ model \cite{Teitel1983} can be achieved through different values of $\alpha$ which reflects the deep connection between the frustrated $XY$ model and the spin-imbalanced Hofstadter-Hubbard model.

\begin{figure}
\centering
\includegraphics[width=\columnwidth]{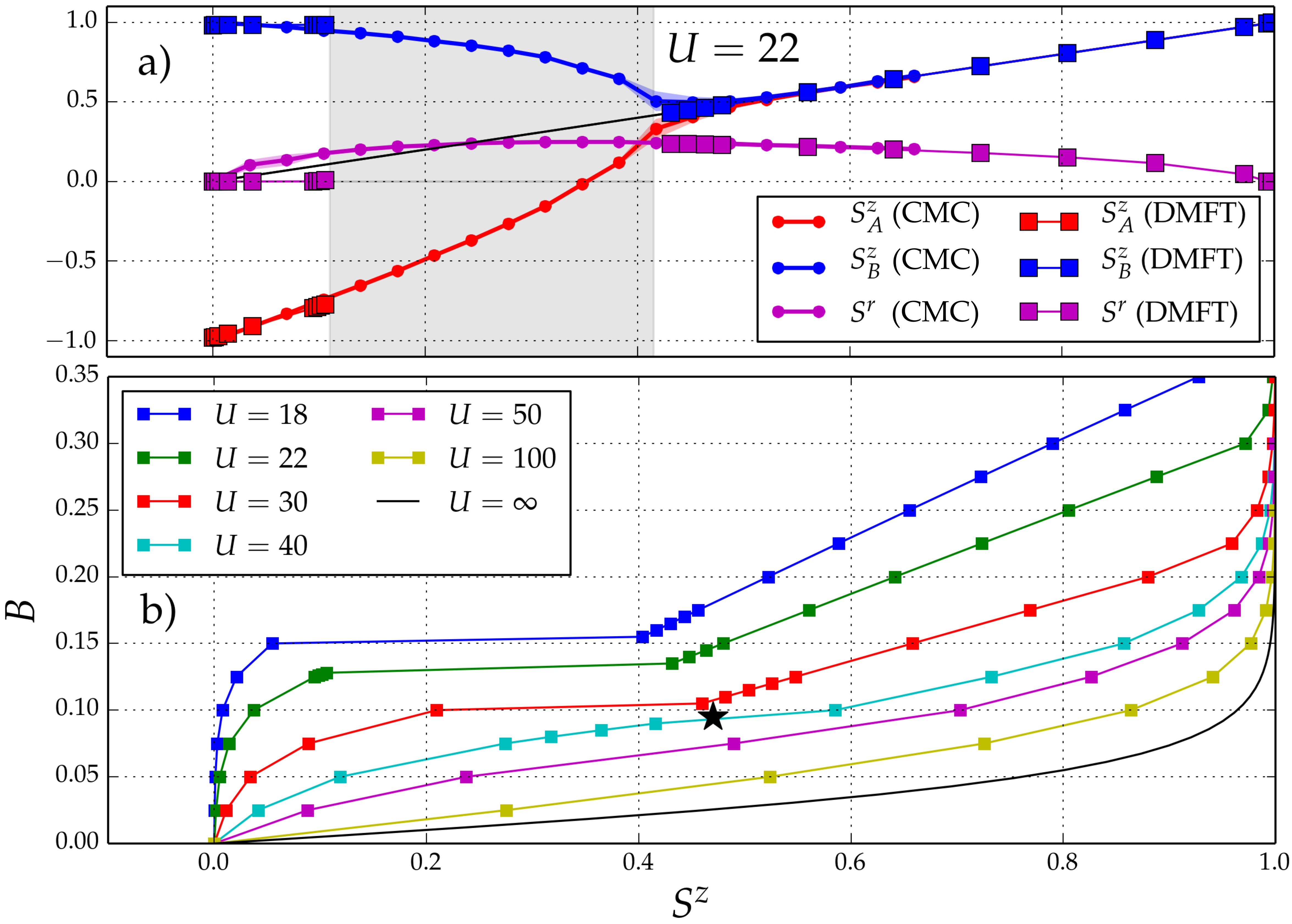}
\caption{a) Comparison between CMC  and DMFT calculations for interaction strength $U=22$. The shaded region is inaccessible for DMFT. b) $S^z$ as function of the applied Zeeman field $B$ from DMFT calculations for different interaction strengths. The black star denotes the critical end-point at which the first-order phase transition from FiM to TM vanishes.}
\label{dmftPre}
\end{figure}

The above considerations have been made in the regime of large interactions, such that only terms quadratic in the hopping energy contribute. We now want to investigate the effect of finite interactions, i.e. quantum fluctuations of the charge degrees of freedom are now present. To this end we make use of real-space DMFT with an exact diagonalization impurity solver with three bath sites and at finite inverse temperature $\beta=20$. DMFT is formulated in the grand-canonical ensemble such that fixing a constant filling can become computationally costly, especially with additional spin-imbalance. From the CMC results we find that fixing an exact imbalance $S^z$ is not necessary to observe the TM. Thus we perform DMFT calculations with fixed Zeeman field $B$. The half-filling condition is satisfied by $\mu_\sigma=U/2$ due to the particle-hole symmetry of the Hamiltonian \eqref{ham}. This is also true for the spin-imbalanced case $\mu_\sigma=U/2\pm B$. As initial guess for the self-energy $\Sigma_{\bm{j},\alpha\beta}$ in Hartree-Fock approximation $\Sigma_{\bm{j},\alpha\beta}^\text{HF}=\langle c_{\bm{j},\alpha}^\dag c_{\bm{j},\beta}\rangle$ we use the spin state from the CMC calculations  for $S^z=1/3$ which is shown in Fig.~\ref{result1_4}b).

In Fig.~\ref{dmftPre}a) we compare DMFT results for interaction strength $U=22$ with CMC results from Fig.~\ref{result1_4}c). We observe that for the regime $S^z\gtrsim0.4$ the agreement between the two methods is perfect, which corresponds to the phase of finite TM and full polarization of the $S^z$-components. Also the case $S^z=0$ yields the AFM result for both theories. However, we find deviations in the regime $0<S^z\lesssim0.4$. Here, the CMC results show the TM with an underlying checkerboard structure of the $S^z$-components as e.g. shown in Fig.~\ref{result1_4}b). In contrast, DMFT results do not exhibit finite $S^r$ in that regime. For $S^z\lesssim0.1$ in DMFT $S^z_B$ remains constantly 1, while $S^z_A$ linearly increases with increasing $S^z$ from -1 with a slope of 2. This corresponds to a ferrimagnetic phase (FiM). Furthermore the shaded regime $0.1\lesssim S^z\lesssim0.4$ seems to be inaccessible for DMFT. This is studied in more detail in Fig.~\ref{dmftPre}b), which shows DMFT results of $S^z$ as a function of $B$ for different interaction strengths $U$. For $U=18,22,30$ we observe a first-order phase transition at $B\approx0.1,0.125,0.15$  between the FiM and the TM phase. With increasing interaction strength the first-order jump shrinks and finally closes at the critical end-point represented by a black star in Fig.~\ref{dmftPre}b). The system then features a crossover from the AFM to the FM phase for very strong interactions $U\gtrsim40$. In this regime the superexchange interaction becomes comparable to the Zeeman field $J=t^2/U\sim B$. For infinite Hubbard interaction strength $U$ the superexchange is completely suppressed at finite values of $B$ and the only remaining term in the Hamiltonian is $-B\sum_{\bm{j}}S_{\bm{j}}^z$. The value of $S^z$ can then be computed from the fermionic distribution as

\begin{equation}
S^z=\frac{1}{1+\exp\left(-\beta B\right)} - \frac{1}{1+\exp\left(\beta B\right)},
\end{equation}
which is shown as black line in Fig.~\ref{dmftPre}b). We observe that this line is indeed approached by the DMFT results with increasing interaction strength.

\begin{figure}
\centering
\includegraphics[width=\columnwidth]{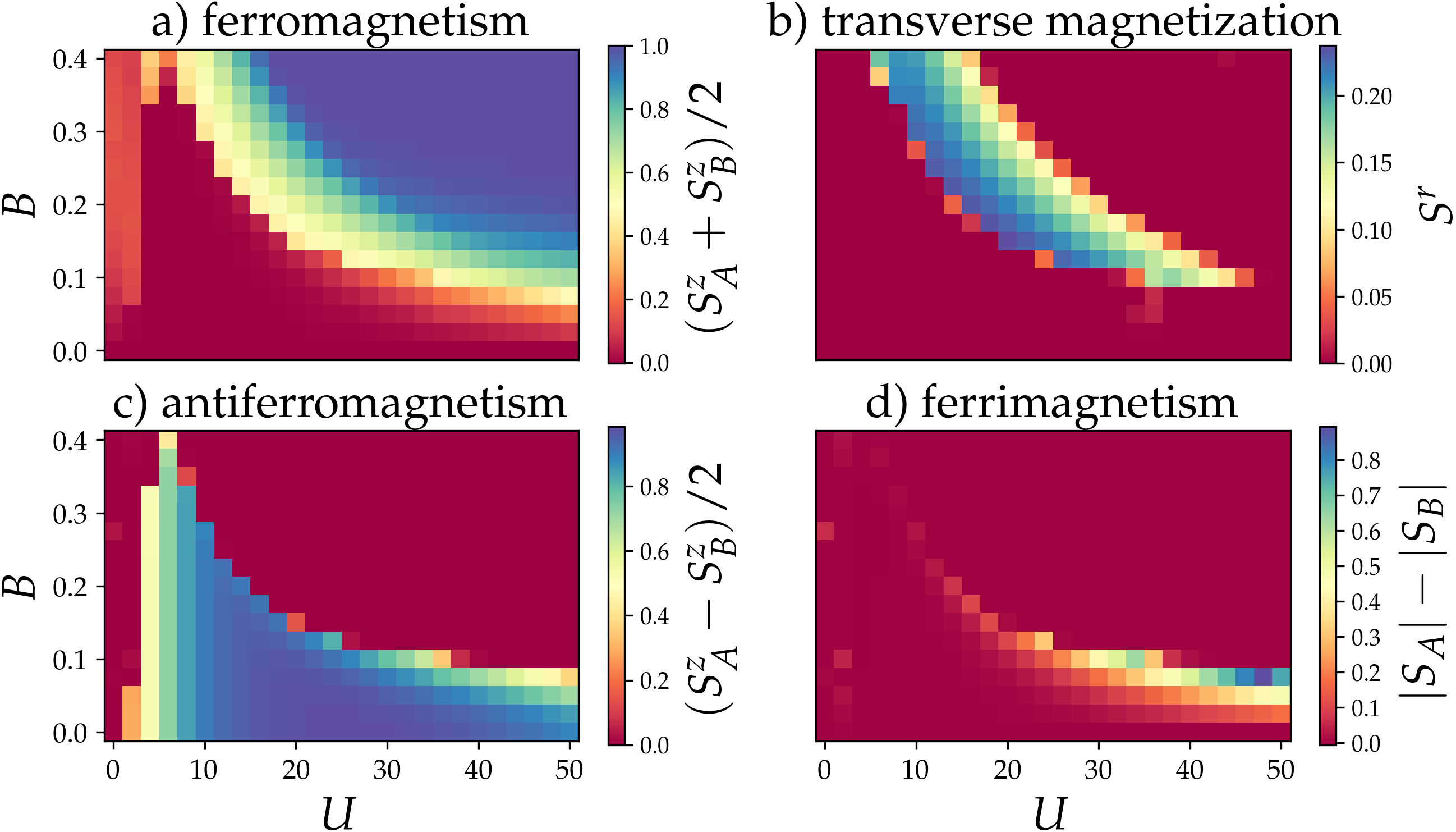}
\caption{Order parameters of the spin-imbalanced Hofstadter-Hubbard model obtained from DMFT calculations: a) ferromagnetic, b) transverse magnetization, c) anti-ferromagnetic, and d) ferrimagnetic}
\label{dmft}
\end{figure}

In Fig.~\ref{dmft}a) we show the phase diagrams of the spin-imbalanced Hofstadter-Hubbard model in the $U$-$B$-parameter space, obtained from the order parameters for FM $(S^z_A+S^z_B)/2$, TM $S^r$ defined in Eq.~\eqref{trans}, AFM $(S^z_A-S^z_B)/2$, and FiM $|S_A|-|S_B|$. The latter is a measure of the difference in length of the spins of the two sublattices $A$ and $B$. 

For experimental probing of the spin structure we propose measurements with a quantum gas microscope similar to \cite{Mazurenko2017}. Here the $S^z_{\bm{j}}$ component of a spin at lattice site $\bm{j}$ can be measured by selective removal of one spin state before imaging via Stern-Gerlach. Other components can be measured by first applying a global spin rotation \cite{Brown2017}. Since there is still rotational symmetry of the system around the $S^z$-direction, this should be broken beforehand by applying a small transverse field in $S^x$-direction. A mapping to a ferromagnetic order through coherent spin manipulation similar to Ref.~\cite{Wurz2018} could also be candidate for a measurement. Since fermionic superexchange temperatures are low and challenging to achieve experimentally and the effect also occurs on a classical level, the TM could also be observed through a mapping to local phases of Bose-Einstein condensates in one-dimensional tubes of an optical lattice \cite{Eckardt2010,Struck2011,Struck2013}. This would then require only temperatures below the critical temperature for Bose-Einstein condensation.

In conclusion, we have studied the spin-imbalanced Hofstadter-Hubbard model with a flux $\alpha=1/4$. In the limit of strong interactions, such that charge degrees of freedom are frozen, we find a spin structure emerging from spin frustration with a finite net magnetization in the transverse direction, i.e. a transverse magnetization effect. We investigate the stability of this intriguing phase against quantum fluctuations and obtain a rich phase diagram. Possible experimental realizations for cold atom setups are discussed. In the present manuscript we use single-site DMFT, i.e., non-local corrections to the selfenergy are neglected. A first extension would be two-site cluster DMFT, which includes corrections to the hopping  matrix element. We can think of two different scenarios for what could happen when including these corrections. Either the transverse magnetization phase is stable since the resulting transverse magnetization is large, i.e., of the same order of magnitude as the longitudinal magnetization imposed by the spin-imbalance, and quantum fluctuations cannot fully destroy it. On the other hand, if additional fluctuations are strong enough, the induced geometric frustration could lead to a break-down of magnetic order. It is thus conceivable that this system is a candidate for a quantum spin liquid.

\begin{acknowledgements}
The authors would like to thank Daniel Cocks, Karyn Le Hur, Jaromir Panas and Christof Weitenberg for enlightening discussions.
This work was supported by the Deutsche Forschungsgemeinschaft (DFG, German Research Foundation) under Project No. 277974659 via Research Unit FOR 2414. This work was also supported by the Deutsche Forschungsgemeinschaft (DFG) via the high-performance computing center LOEWE-CSC. 
\end{acknowledgements}

\bibliographystyle{apsrev4-1}
%

\end{document}